\documentclass[10pt,twocolumn,letterpaper]{article}

\usepackage{cvpr}              

\usepackage{graphicx}
\usepackage{amsmath}
\usepackage{amssymb}
\usepackage{booktabs}
\usepackage{float}
\usepackage{stfloats}
\usepackage{caption}

\usepackage{multirow}
\usepackage[symbol]{footmisc}
\usepackage[accsupp]{axessibility}  

\renewcommand{\thefootnote}{\fnsymbol{footnote}}

\usepackage[pagebackref,breaklinks,colorlinks]{hyperref}

\usepackage[capitalize]{cleveref}
\crefname{section}{Sec.}{Secs.}
\Crefname{section}{Section}{Sections}
\Crefname{table}{Table}{Tables}
\crefname{table}{Tab.}{Tabs.}

\begin{document}

\title{Improving visual image reconstruction from human brain activity using latent diffusion models via multiple decoded inputs}

\author{
Yu Takagi${}^\text{1,2}$\footnotemark \quad\quad Shinji Nishimoto${}^\text{1,2}$ \vspace{0.2cm} \\
${}^\text{1}$Graduate School of Frontier Biosciences, Osaka University, Japan \vspace{0.2cm}\\
${}^\text{2}$CiNet, NICT, Japan\\
{\tt\small \{takagi.yuu.fbs,nishimoto.shinji.fbs\}@osaka-u.ac.jp}
}

\maketitle

\footnotetext{* Corresponding author}

\renewcommand{\thefootnote}{\arabic{footnote}}

\begin{abstract}
The integration of deep learning and neuroscience has been advancing rapidly, which has led to improvements in the analysis of brain activity and the understanding of deep learning models from a neuroscientific perspective.
The reconstruction of visual experience from human brain activity is an area that has particularly benefited: the use of deep learning models trained on large amounts of natural images has greatly improved its quality, and approaches that combine the diverse information contained in visual experiences have proliferated rapidly in recent years.
In this technical paper, by taking advantage of the simple and generic framework that we proposed \cite{takagi2023high}, we examine the extent to which various additional decoding techniques affect the performance of visual experience reconstruction.
Specifically, we combined our earlier work \cite{takagi2023high} with the following three techniques: using decoded text from brain activity, nonlinear optimization for structural image reconstruction, and using decoded depth information from brain activity. We confirmed that these techniques contributed to improving accuracy over the baseline. We also discuss what researchers should consider when performing visual reconstruction using deep generative models trained on large datasets.
Please check our webpage at \href{https://sites.google.com/view/stablediffusion-with-brain/}{https://sites.google.com/view/stablediffusion-with-brain/}. Code is also available at \href{https://github.com/yu-takagi/StableDiffusionReconstruction}{https://github.com/yu-takagi/StableDiffusionReconstruction}.

\end{abstract}

\section{Introduction}
\label{sec:intro}
The integration of deep learning into neuroscience is advancing rapidly. Visual image reconstruction from human brain activity, measured, for example, using functional magnetic resonance imaging (fMRI), is an area that has particularly benefited from deep learning models trained on large datasets of natural images \cite{yamins2014performance,horikawa2017generic,kietzmann2019recurrence,gucclu2015deep,groen2018distinct,wen2018neural,kell2018task,koumura2019cascaded,schrimpf2021neural,goldstein2022shared,caucheteux2022brains,schmitt2021predicting,sun2023contrast}. In studies on the reconstruction of visual experiences, approaches that combine diverse information contained in visual experiences have increased rapidly in recent years \cite{takagi2023high,zeng2023controllable,chen2023cinematic,ozcelik2023brain,lin2022mind,lu2023minddiffuser,scotti2023reconstructing,koide2023mental,gu2022decoding,chen2023seeing,ferrante2023brain,liu2023brainclip}. The availability of large publicly available fMRI datasets with richly annotated stimuli has contributed to this, for example,  the Natural Scenes Dataset (NSD) \cite{Allen2022}, in addition to multimodal deep generative models that have been pre-trained on large datasets, for example, Stable Diffusion \cite{rombach2022high}. Numerous techniques have been proposed by these ongoing works to improve the performance of visual experience reconstruction, including using decoded textual information predicted from brain activity, nonlinear optimization for image reconstruction, and using decoded depth information predicted from brain activity.

The extent to which each technique contributes to improving the performance of visual experience reconstruction is an interesting issue for further research. Among the many approaches, our proposed framework can be characterized by its simplicity and generality \cite{takagi2023high} (Figure \ref{fig:schematic}\textbf{a}). Using these features, our framework is flexible enough to incorporate various techniques and can provide useful information for examining to what extent each technique affects reconstruction performance.

The purpose of this technical paper is to provide quantitative references by combining our previous method \cite{takagi2023high} and additional techniques (\ref{subsec:caption}, \ref{subsec:gan}, and \ref{subsec:depth}). We also provide additional control analyses regarding exercising caution when performing visual reconstruction using deep generative models trained on large datasets (\ref{subsec:control}).

\section{Results}
\label{sec:results}
\begin{figure*}[ht]
  \centering
   \includegraphics[width=1\linewidth]{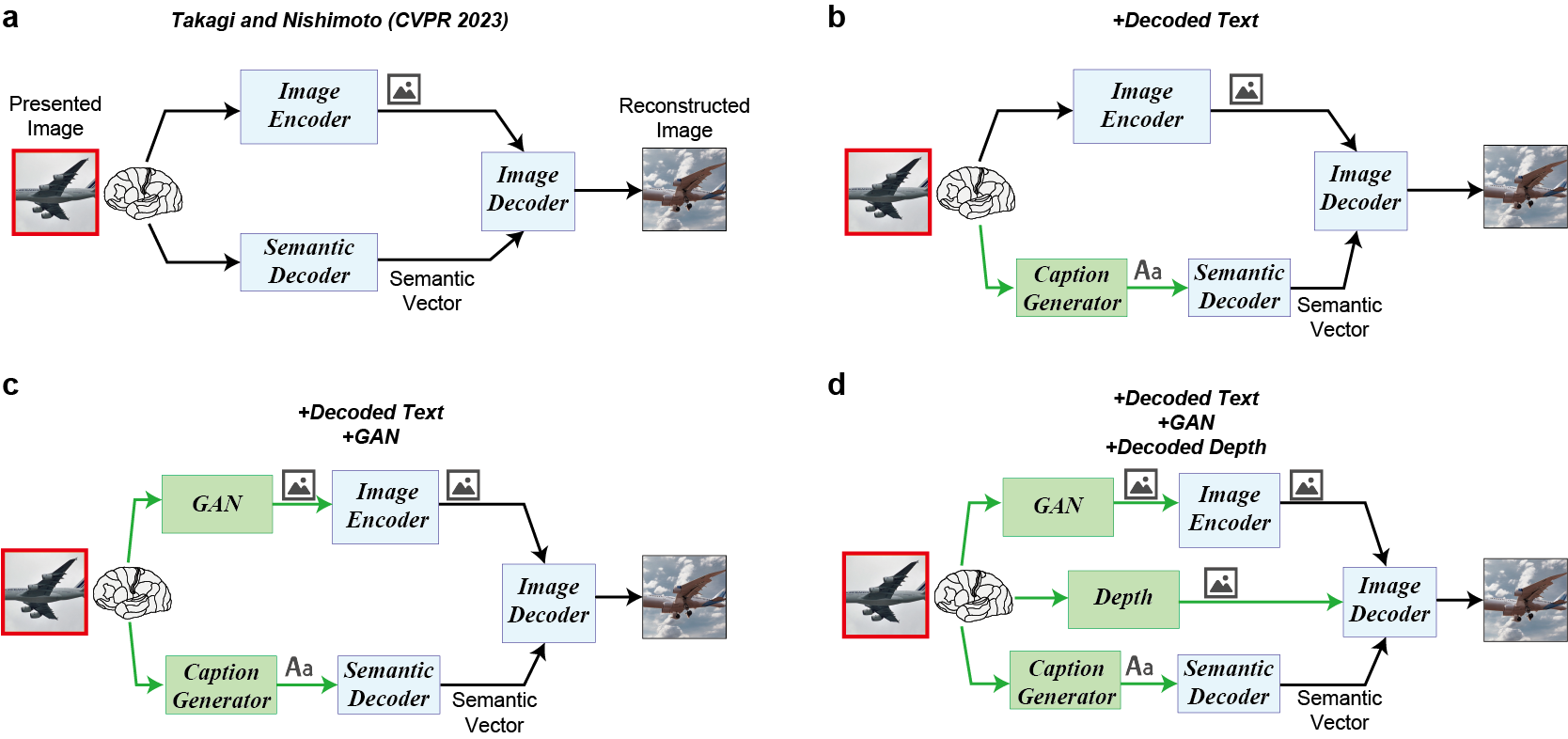}
   \caption{Schematic of the approach in this paper, including (\textbf{a}) our earlier work \cite{takagi2023high} and (\textbf{b}--\textbf{d}) with three additional techniques.}
   \label{fig:schematic}
\end{figure*}

\begin{figure}[ht]
  \centering
   \includegraphics[width=1\linewidth]{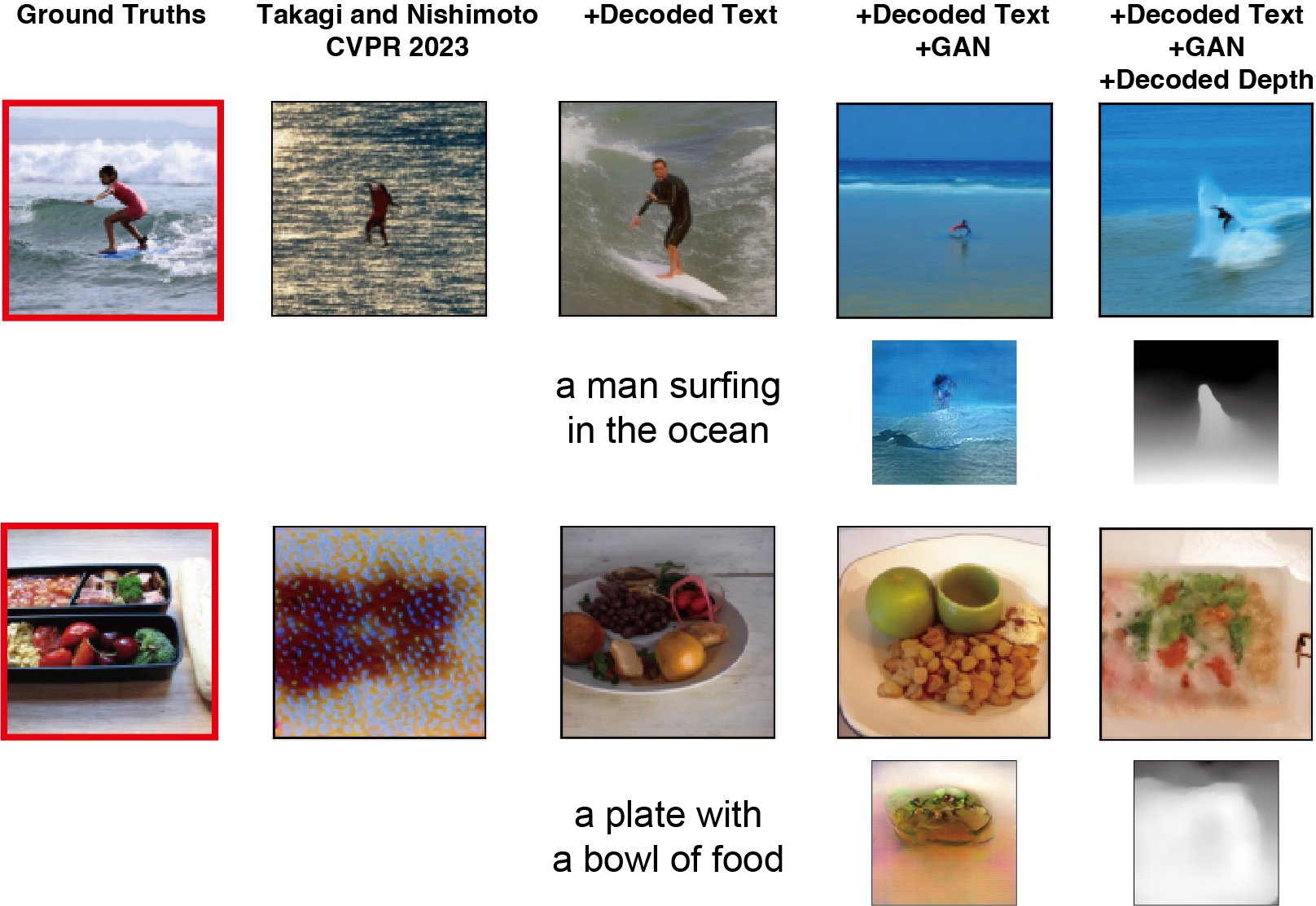}
   \caption{Examples of the presented (red box) and reconstructed images using various methods. The decoded text, image produced by GAN, and decoded depth are shown below the reconstructed image.}
   \label{fig:example}
\end{figure}

\begin{figure*}[ht]
  \centering
   \includegraphics[width=1\linewidth]{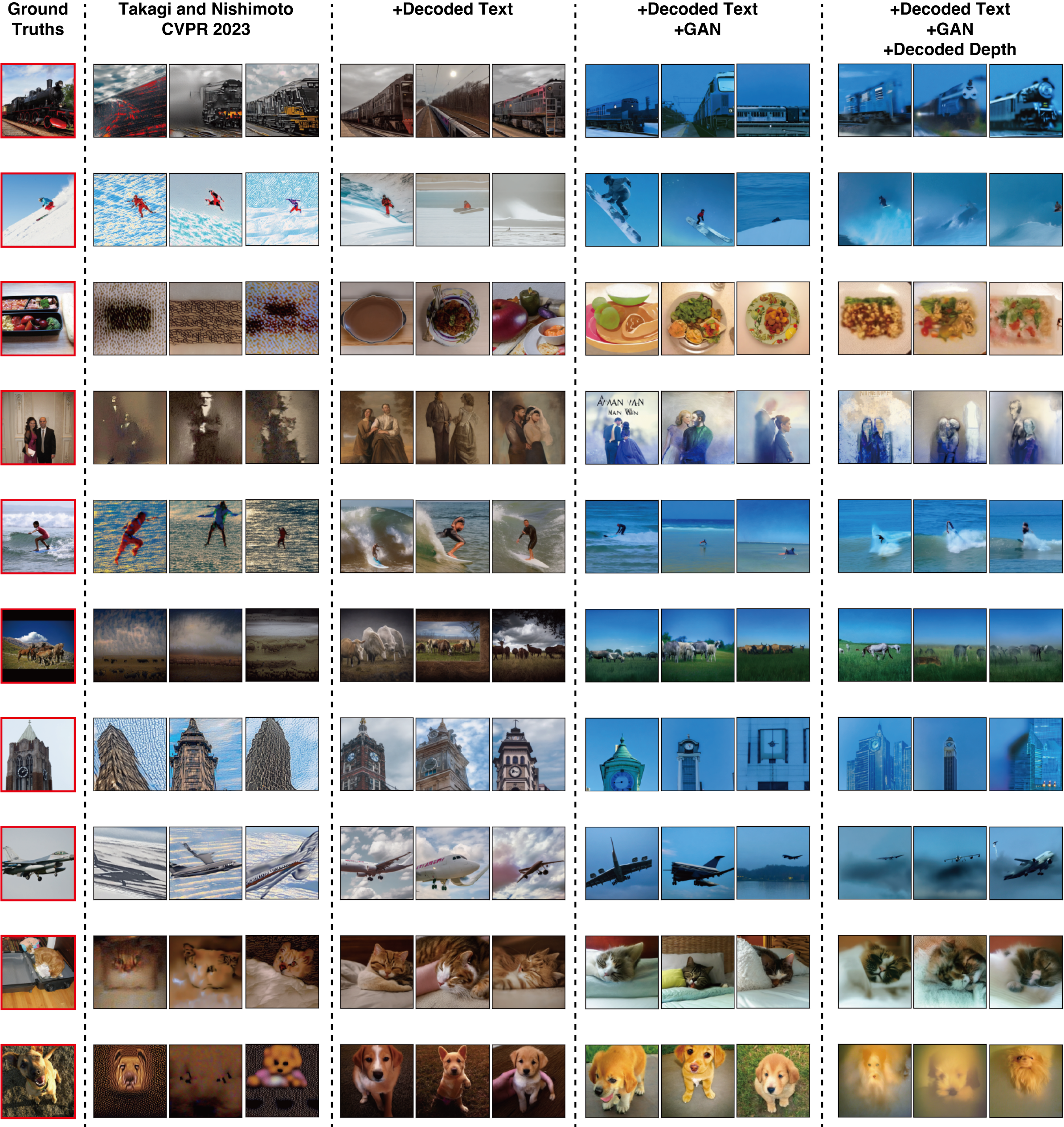}
   \caption{Perceived images (leftmost, red box) and examples of reconstructed images from the brain activity of a single subject (\textit{subj01}) using various decoding methods (separated by dashed lines). For each method, three generated images from different stochastic noise were chosen randomly.}
   \label{fig:manyexample}
\end{figure*}

All analyses in this paper build on our previous paper \cite{takagi2023high}. For more details on the method, we recommend that the reader refers to our previous paper.
Figure \ref{fig:schematic} shows a schematic of how we combine various techniques with our prior work. Figure \ref{fig:example} shows an example of decoding results from our prior work and the three additional techniques for an example subject (\textit{subj01}). 
Figure \ref{fig:manyexample} shows more examples from a single subject (\textit{subj01}). For each original image and for each method, we generated five images with different stochastic noise and selected three images randomly.

Table \ref{table:accuracy} shows the quantitative evaluation results for all methods. We described the detailed quantification procedures in the original paper \cite{takagi2023high}. Briefly, to take into account the variability of image generation across different stochastic noise, we generated five images with different noise for each test sample. Then we calculated the average quality of the five images via the two-way identification accuracy \cite{takagi2023high} using various image features.

We found that these techniques generally improved accuracy over the baseline. However, we also found that each technique did not necessarily improve all measures for all subjects. The degree of improvement over the baseline varied from subject to subject and from measure to measure. 
We describe the details of each method below.

\subsection{Using decoded captions from brain activity}
\label{subsec:caption}
In our earlier work \cite{takagi2023high}, we predicted semantic latent representations of the caption associated with each image (CLIP embedding) from brain activity and used them as input into Stable Diffusion. However, the reconstructed images were blurred compared with the realistic images that Stable Diffusion would normally output. Interestingly, although the images generated from brain activity were blurred, they retained the semantic content of the original images. This suggests that the semantic conditioning of Stable Diffusion is vulnerable to noise associated with estimation from brain activity.

Based on the above observation, we replaced the problem of predicting semantic latent representations of images from brain activity with the problem of estimating image captions from brain activity (Figure \ref{fig:schematic}\textbf{b}). We used an image captioning model \cite{li2022blip} called BLIP to generate captions from brain activity. BLIP extracts Vision Transformer (ViT) features from the input image. Then we generated a full sentence using BERT \cite{devlin2018bert} via cross-attention to the ViT features. We used the officially released base version of the model trained on 129M images \footnote{BLIP w/ ViT-B and the CapFilt-L model in \url{https://github.com/salesforce/BLIP}.}. We used default values for all hyperparameters, except for the penalty of repetition, which we set to 1.5. To decode a caption from brain activity, we predicted ViT features from brain activity. We used early, ventral, midventral, midlateral, lateral, and parietal visual regions included in the \textit{streams} atlas. We estimated the weights of the prediction model from training data using L2 regularized linear regression and then applied the model to test data. We used the same visual regions and analysis framework for the subsequent generative adversarial network (GAN) and depth decoding. Please see our code for the detailed implementation. 

We confirmed that the proposed method generated captions that well represented the human visual experience of the subject. Quantitative evaluation also showed that captions generated from the original image explained the original image better than captions assigned to other randomly selected images (result not shown). We also confirmed that this approach improved reconstruction performance, as shown in Table \ref{table:accuracy}. This improvement has been reported in other ongoing studies \cite{ferrante2023brain}, including ours\footnote{ \url{https://www.anlp.jp/proceedings/annual_meeting/2023/pdf_dir/B8-2.pdf}.}.


\subsection{Using nonlinear optimization with GAN}
\label{subsec:gan}
In our earlier work \cite{takagi2023high}, we reconstructed a visual image by predicting latent visual representations of the bottleneck layer in the variational autoencoder from brain activity. This method is simple and straightforward because it only requires the estimation of a linear model to predict low-dimensional representations in Stable Diffusion.

Using the modular nature of our previous framework \cite{takagi2023high}, we can also incorporate more complex decoding procedures flexibly. We examined such an extension using GAN-based decoding as described in \cite{Shen2019}. We first decoded brain activity using the GAN-based algorithm\footnote{\url{https://github.com/KamitaniLab/brain-decoding-cookbook-public/tree/main/reconstruction}} and then fed the decoded results into the image encoder block to generate images (Figure \ref{fig:schematic}\textbf{c}).

As expected, reconstruction performance tended to increase, particularly when the reconstructed images were evaluated with low-level latent features (see  Table \ref{table:accuracy}). Several ongoing works have been proposed that use more complex optimization than we have used in the present paper. However, note that such methods usually require complex and computationally intensive optimization, which is a non-trivial task.


\subsection{Using visual depth information}
\label{subsec:depth}
In our earlier work \cite{takagi2023high}, we integrated low-level visual and high-level semantic information predicted from brain activity. However, human visual experience contains more diverse information. Specifically, depth information, for example, is processed in the human visual cortex \cite{lescroart2019human}. To demonstrate the further flexibility of our framework, we separately estimated depth information and then performed visual reconstruction that incorporated depth information (Figure \ref{fig:schematic}\textbf{d}).

We used Stable Diffusion 2.0 to reconstruct visual experiences conditioned on both semantic and depth information. Note that, we confirmed that there is no change in reconstruction accuracy between Stable Diffusion 2.0 and Stable Diffusion 1.4 (also see the description in Section 4.). To predict depth information from brain activity, we first extracted the latent representation of the DPT model implemented by Huggingface \cite{Ranftl2021,}\footnote{Intel/dpt-large model in \url{https://huggingface.co/docs/transformers/main/model_doc/dpt}} for each image. Then we reconstructed the depth image estimated from brain activity. 

We confirmed that depth information improved the quantitative evaluation (CLIP(Late)) across subjects. Qualitative analysis confirmed that when depth information was estimated appropriately, the generated images had a fairly stable appearance across various generations (see Figure \ref{fig:manyexample}). By contrast, conditioning by depth information strongly influenced the generated image; hence, there was a significant negative effect when the depth estimation was wrong.

\subsection{Control analysis}
\label{subsec:control}
After publishing our earlier work \cite{takagi2023high}, we noticed an overlap between the data used to train Stable Diffusion (v1.4) and the images displayed in fMRI. To investigate whether such an overlap affects reconstruction performance, we performed the following control analysis.

First, we discovered that approximately 7\% of the images used in the test data were present in LAION-5B\footnote{Among 35 example images shown in the paper, three were included in LAION-5B (Figure 3, line 1 [identical to Sup Fig. B4, column 2, line 7 and Sup Fig. B5, column 1, line 4]; Figure 4, line 3; and Sup. Fig. B4, column 2, line 8).}, which Stable Diffusion used for training. Based on this observation, we conducted a quantitative evaluation again and excluded overlapping images. We found no change in the quantitative evaluation (identification accuracy = 74.3 ± 1.7\%/74.3 ± 1.6\% [previous/new] when using Inception v3. There were no differences in other quantitative measures when we used CLIP and AlexNet).

To further investigate whether the overlap in the text encoding process might impact the results, we performed the same quantitative analysis using Stable Diffusion v2.0 trained with OpenCLIP (trained on LAION-5B) instead of CLIP (trained on MS COCO, which is the source of NSD stimuli). We again found no change in the quantitative evaluation (identification accuracy = 74.3 ± 1.7\%/74.5 ± 2.7\% [previous/new] when using Inception v3. There were no differences in other quantitative measures using CLIP and AlexNet).

These results indicate that potential image leakage between Stable Diffusion and NSD did not affect the conclusions of our prior work \footnote{The method, image list, and results of the investigation of the overlap between NSD and LAION-5B are available at \url{https://drive.google.com/drive/folders/1J9yn7I2ql2-1kmG4MLp__7HsJz5tQV_c?usp=sharing}.}.

\begin{table*}[!ht]
\centering
\small
\begin{tabular}{lllllllll}
                         &          & \multicolumn{1}{l}{CLIP(Early)} & \multicolumn{1}{l}{CLIP(Middle)} & \multicolumn{1}{l}{CLIP(Late)} & \multicolumn{1}{l}{CNN(Early)} & \multicolumn{1}{l}{CNN(Middle)} & \multicolumn{1}{l}{CNN(Late)} & \multicolumn{1}{l}{Inception-v3} \\ \hline\hline
\multirow{4}{*}{subj01}                   & \cite{takagi2023high} & 0.795                     & 0.732                      & 0.774                    & 0.851                        & 0.831                         & 0.830                         & 0.759                         \\
                         & +\textbf{T}       & \textbf{0.922}                     & 0.836                      & 0.859                    & 0.901                        & 0.934                         & 0.926                         & 0.877                         \\
                         & +\textbf{T}+\textbf{G}     & 0.910                     & \textbf{0.848}                      & 0.870                    & \textbf{0.922}                        & \textbf{0.949}                         & \textbf{0.938}                         & \textbf{0.879}                         \\
                         & +\textbf{T}+\textbf{G}+\textbf{D}   & 0.901                     & 0.829                      & \textbf{0.877}                    & 0.905                        & 0.944                         & 0.917                         & 0.856                         \\ \hline
\multirow{4}{*}{subj02}  & \cite{takagi2023high} & 0.772                     & 0.709                      & 0.768                    & 0.830                        & 0.819                         & 0.815                         & 0.725                         \\
                         & +\textbf{T}       & \textbf{0.904}                     & 0.818                      & 0.842                    & 0.873                        & 0.922                         & 0.923                         & 0.854                         \\
                         & +\textbf{T}+\textbf{G}     & 0.889                     & \textbf{0.834}                      & 0.853                    & \textbf{0.912}                        & \textbf{0.944}                         & \textbf{0.933}                         & \textbf{0.867}                         \\
                         & +\textbf{T}+\textbf{G}+\textbf{D}   & 0.885                     & 0.821                      & \textbf{0.869}                    & 0.892                        & 0.934                         & 0.905                         & 0.848                         \\ \hline
\multirow{4}{*}{subj05}  & \cite{takagi2023high} & 0.742                     & 0.707                      & 0.778                    & 0.800                        & 0.817                         & 0.825                         & 0.761                         \\
                         & +\textbf{T}       & \textbf{0.874}                     & \textbf{0.808}                      & 0.846                    & \textbf{0.834}                        & \textbf{0.914}                         & \textbf{0.922}                         & 0.866                         \\
                         & +\textbf{T}+\textbf{G}     & 0.802                     & 0.779                      & 0.852                    & 0.809                        & 0.905                         & 0.920                         & \textbf{0.867}                         \\
                         & +\textbf{T}+\textbf{G}+\textbf{D}   & 0.788                     & 0.753                      & \textbf{0.861}                    & 0.791                        & 0.897                         & 0.885                         & 0.842                         \\ \hline
\multirow{4}{*}{subj07}  & \cite{takagi2023high} & 0.726                     & 0.693                      & 0.753                    & 0.775                        & 0.793                         & 0.801                         & 0.726                         \\
                         & +\textbf{T}       & 0.839                     & 0.775                      & 0.819                    & 0.799                        & 0.885                         & 0.890                         & 0.835                         \\
                         & +\textbf{T}+\textbf{G}     & \textbf{0.861}                     & \textbf{0.814}                      & 0.836                    & \textbf{0.875}                        & \textbf{0.919}                         & \textbf{0.912}                         & \textbf{0.843}                         \\
                         & +\textbf{T}+\textbf{G}+\textbf{D}   & 0.845                     & 0.781                      & \textbf{0.858}                    & 0.846                        & 0.907                         & 0.885                         & 0.825                         \\ \hline\hline
\multirow{4}{*}{\begin{tabular}{ll}
    Average ±  \\
    s.t.d.
  \end{tabular}} & \cite{takagi2023high} & 0.759 ± 0.027             & 0.710 ± 0.014              & 0.768 ± 0.010            & 0.814 ± 0.029                & 0.815 ± 0.014                 & 0.818 ± 0.011                 & 0.743 ± 0.017                 \\
                         & +\textbf{T}       & \textbf{0.885 ± 0.032}             & 0.809 ± 0.022              & 0.841 ± 0.014            & 0.852 ± 0.039                & 0.914 ± 0.018                 & 0.915 ± 0.014                 & 0.858 ± 0.016                 \\
                         & +\textbf{T}+\textbf{G}     & 0.866 ± 0.041             & \textbf{0.819 ± 0.026}              & 0.853 ± 0.012            & \textbf{0.880 ± 0.044}                 & \textbf{0.929 ± 0.018}                 & \textbf{0.926 ± 0.010}                 & \textbf{0.864 ± 0.013}                 \\
                         & +\textbf{T}+\textbf{G}+\textbf{D}   & 0.855 ± 0.043             & 0.796 ± 0.031              & \textbf{0.866 ± 0.007}            & 0.858 ± 0.045                & 0.921 ± 0.019                 & 0.898 ± 0.014                 & 0.843 ± 0.011                 \\ \hline\hline
\end{tabular}
\caption{Two-way identification accuracy. \textbf{T} denotes a method that uses decoded text predicted from brain activity; \textbf{G} denotes a method that uses GAN to input the image into Stable Diffusion; and \textbf{D} denotes a method that uses decoded depth information from brain activity.}
\label{table:accuracy}
\end{table*}

\section{Discussion}
\label{sec:discussion}
We took advantage of the simplicity and generality of our earlier work \cite{takagi2023high} to examine how various techniques affect the performance of visual experience reconstruction. We found that these techniques improve accuracy in general over the baseline. However, we also found that each technique did not necessarily improve all measures for all subjects. The degree of improvement from the baseline varied from subject to subject and from measure to measure. We believe that such information will be useful for conducting future studies with quantitative evaluation and comparisons.


\section*{Acknowledgements}
We would like to thank the Computer Vision and Learning research group at Ludwig Maximilian University of Munich for providing the code and models for Stable Diffusion, Stability AI and Runway for supporting the training of Stable Diffusion, and NSD for providing the neuroimaging dataset. We would like to thank Tim Kietzmann for useful discussions. We used the Himalaya library \cite{laTour2022} for the main analysis. YT was supported by JSPS KAKENHI (19H05725). SN was supported by MEXT/JSPS KAKENHI JP18H05522, in addition to JST CREST JPMJCR18A5 and ERATO JPMJER1801. 

{\small
\bibliographystyle{ieee_fullname}
\bibliography{ref}
}

\end{document}